\documentclass[citeautoscript,floatfix,aps,prl,onecolumn,superscriptaddress]{revtex4-1}
\usepackage{graphicx}
\usepackage{amsmath}
\usepackage{amssymb}
\usepackage{dcolumn}
\usepackage[version=4]{mhchem}
\usepackage{latexsym}
\usepackage{rotating}
\usepackage{epstopdf}
\usepackage[usenames,dvipsnames]{xcolor}
\usepackage{float}
\usepackage{soul}
\usepackage{epsfig}
\usepackage{psfrag}
\usepackage{natbib}
\usepackage{bm}
\usepackage{eucal}
\usepackage{mathrsfs}
\usepackage{braket}
\usepackage{enumerate}
\usepackage{longtable}
\usepackage{bm}
\usepackage{hyperref}
\usepackage{amsfonts}
\usepackage{dcolumn}
\usepackage{bm}
\usepackage{changes}

\newcommand{\be}{\begin{equation}}
\newcommand{\ee}{\end{equation}}
\newcommand{\bn}{\begin{eqnarray}}
\newcommand{\en}{\end{eqnarray}}

\def\x2y2{{x^2-y^2}}

\usepackage{color} 

\usepackage{hyperref}
\hypersetup{
colorlinks=true,final=true,
        linkcolor=red,
        citecolor=blue,
        filecolor=blue,
        urlcolor=blue,
}
\begin{document}

\title{Excitons in Bulk and Layered Chromium Tri-Halides: From Frenkel to the Wannier-Mott Limit}
\author{Swagata Acharya}
\affiliation{Institute for Molecules and Materials, Radboud University, {NL-}6525 AJ Nijmegen, The Netherlands}	
\email{swagata.acharya@ru.nl}
\author{Dimitar Pashov}
\affiliation{ King's College London, Theory and Simulation of Condensed Matter,
	The Strand, WC2R 2LS London, UK}
\author{Alexander N. Rudenko}
\affiliation{Institute for Molecules and Materials, Radboud University, {NL-}6525 AJ Nijmegen, The Netherlands}
\author{Malte R\"{o}sner}	
\affiliation{Institute for Molecules and Materials, Radboud University, {NL-}6525 AJ Nijmegen, The Netherlands}
\author{Mark van Schilfgaarde}
\affiliation{National Renewable Energy Laboratory, Golden, CO 80401, USA}	
\affiliation{ King's College London, Theory and Simulation of Condensed Matter,
	The Strand, WC2R 2LS London, UK}
\author{Mikhail I. Katsnelson}
\affiliation{Institute for Molecules and Materials, Radboud University, {NL-}6525 AJ Nijmegen, The Netherlands}


\begin{abstract}

Excitons with large binding energies $\sim$2-3 eV in CrX$_{3}$ are historically characterized as being localized
(Frenkel) excitons that emerge from the atomic $d{-}d$ transitions between the Cr-3$d$-$t_{2g}$ and $e_{g}$
orbitals. The argument has gathered strength in recent years as the excitons in recently made monolayers are found at
almost the same energies as the bulk. The Laporte rule, which restricts such parity forbidden atomic transitions, can
relax if, at least, one element is present: spin-orbit coupling, odd-parity phonons or Jahn-Teller distortion.  While
what can be classified as a purely Frenkel exciton is a matter of definition, we show using an advanced first principles
parameter-free approach that these excitons in CrX$_{3}$, in both its bulk and monolayer variants, have band-origin and
do not require the relaxation of Laporte rule as a fundamental principle. We show that, the character
  of these excitons is mostly determined by the Cr-$d$ orbital manifold, nevertheless,
they appear only as a consequence of X-p states hybridizing with the Cr-$d$. The hybridization enhances as the halogen atom becomes heavier, bringing the X-$p$
states closer to the Cr-$d$ states in the sequence Cl{\textrightarrow}Br{\textrightarrow}I, with an attendant increase
in exciton intensity and decrease in binding energy.  By applying a range of different kinds of perturbations, we show that, moderate changes to the
two-particle Hamiltonian that essentially modifies the Cr-$d$-X-$p$ hybridization, can alter both the intensities and
positions of the exciton peaks.  A detailed analysis of several deep lying excitons, with and without strain, reveals
that the exciton is most Frenkel like in CrCl$_{3}$ and acquires mixed Frenkel-Wannier character in
CrI$_{3}$.

\end{abstract}
\maketitle

\section*{Introduction} 

Excitons are charge-neutral excitations, and can be well approximated by the eigenstates of a two-particle
Hamiltonian. An electron-hole pair can form an excitonic bound state as the
Coulomb interaction becomes strong. These two-particle bound states can form deep inside one-particle electronic band
gap in insulators. Usually in systems where the valence and conduction states are mostly dominated by $s$ and $p$
electrons the Coulomb interaction is strongly screened and the excitons become weakly bound. Such excitons are refereed
to as Mott-Wannier excitons~\cite{wannier} due to their delocalized nature in real space with diameters up to few
nanometres. These excitons are heavily studied in the literature and usually their binding energies range between tens
of meV to 500 meV~\cite{mos2,ws2}. The excitons observed in transition metal dichalcogenides~\cite{mos2,ws2},
LiF~\cite{santosh} are prototypical example of Wannier-Mott excitons.  Only the bottom of the conduction band and the
top of the valence band take part in formation of these excitons.  Nevertheless, as systems become more strongly
correlated, for example, in systems with low-energy flat $d$-states, the Coulomb interaction between electron and hole
increases and they can have large binding energies with a diameter approximately that of an atom in an extreme
case. Such strongly localized excitons can have binding energies $\sim$1 eV or more and they are refereed to as Frenkel
excitons~\cite{frenkelorig1,frenkelorig2,frenkel1}. Such excitons are heavily studied in molecular
systems~\cite{frenkelmolecule,frenkelmolecule1,frenkelmolecule2} that are far from the band-limit. In a crystalline
environment where a band picture is applicable, an entire range of bands~\cite{santosh} that contain character of the
orbital where the exciton predominantly resides can contribute to their formation, in strong contrast to the
Wannier-Mott excitons.

Recent observations of excitons with high binding energies in two-dimensional ferromagnetic monolayers of
CrX$_{3}$~\cite{zhang,seyler} have rekindled the investigation of Frenkel excitons in real materials.  In a purely
atomic picture, the low energy manifold of these systems are determined by the Cr-$3d\,t_{2g\uparrow}$ valence orbitals
that contributes to magnetic moment of 3 $\mu_{B}$/Cr atom and the Cr-$3d\,e_{g\uparrow}$ conduction orbitals.  In a
series of recent experimental studies that perform controlled photoluminescence and other optical measurements on these
2D magnets, excitons are observed at $\sim$1 eV~\cite{zhang,seyler} while the electronic band gaps of these systems are
in the range of $\sim$3-5 eV.  These materials are, however, not new and their bulk variants had been studied for long
with the earliest studies dating back more than half a
century~\cite{kamimura1966,grant1968,pollini1970,bermudez1979,nosenzo1984}.  To realise these magnets in their 2D
variant is very recent as ferromagnetism (FM) in a monolayer \ce{CrI3} was first reported in 2017~\cite{huang,klein},
which was followed by observation of FM in CrBr$_{3}$~\cite{zhang,kim}, CrCl$_{3}$~\cite{cai} and many other
compounds~\cite{li,fei,deng,gong,gibert}. Bulk and layered variants of the same compound do have different electronic
band gaps due to differences in the screening environment originating from their effective dimensionality. Going from
bulk to monolayer the electronic screening reduces significantly, leading to a screened Coulomb exchange which is larger than the bulk.  Nevertheless, the observation of the same deep lying excitons at
almost the same energies both in bulk and monolayer (ml-) CrX$_{3}$ is intriguing and suggests that these excitons are
Frenkel-like and, could probably be, even the extreme limit of that where they emerge purely from the atomic $d$-$d$
transitions~\cite{seyler}.

In a recent work~\cite{swagcrx1} we showed that both the electronic band gaps and the halogen components in
the valence band manifold changes significantly in ml-CrX$_{3}$ depending on the nature of the ligand ($X$) atom. There
is significant hybridization of the Cr-$3d$ states with the X-$p$ states and the degree of hybridization increases
significantly in the sequence Cl{\textrightarrow}Br{\textrightarrow}I.  In this work we show that this hybridization
fully controls the exciton intensity, and also to some extent its position.

	\begin{table}[bth]
		\footnotesize
		\begin{tabular}{c|@{\hskip 3pt}ccc@{\hskip 3pt}|@{\hskip 3pt}ccc}
			\hline
			theory & \multicolumn{3}{c}{Monolayer bandgap $E_{g}$ (eV)} & \multicolumn{3}{c}{Bulk bandgap $E_{g}$ (eV)} \cr
			& \ce{CrCl3} & \ce{CrBr3}  & \ce{CrI3} & \ce{CrCl3} & \ce{CrBr3} & \ce{CrI3} \cr
			\hline
			DFT    & 1.51  & 1.30   & 1.06   & 1.38 & 1.21 & 0.9  \cr
			QSGW   & 6.87  & 5.73   & 3.25   &  5.4 & 4.38 &  3.0 \cr
			QS$G\widehat{W}$    & 5.55  & 4.65   & 2.9    & 4.4 & 3.5 & 2.5  \cr
			G$_{0}$W$_{0}$ & 5.47~\cite{molina} & 4.45~\cite{molina}, 3.8~\cite{louie1} & 2.76~\cite{molina} & & & \cr
		\end{tabular}
		\caption{One particle electronic band gap $E_{g}$ at different levels of theory (with spin-orbit coupling). The gap increases from LDA to QS\emph{GW} level. When
			when ladder diagrams are added two-particle interactions via a BSE, $W{\rightarrow}{\widehat{W}}$ and screening is increased.
			This reduces the QS\emph{GW} bandgap by $\sim$20-25\%. Also $E_{g}$ reduces in bulk compared to their monolayer variants by $\sim$20\%. 
                        The last line notes E$_{g}$ values as reported from LDA based single shot G$_{0}$W$_{0}$ calculations performed by different groups.}
		\label{tab:gaps}
	\end{table}

We employ our advanced many-body perturbative Green's function based approach and show how in these materials excitons
with high binding energies are sensitive to fine details of the two-particle Hamiltonian. Following our earlier work~\cite{swagcrx1} we
employ three different levels of theory: the local-density approximation (LDA), quasiparticle self-consistent \emph{GW}
theory~\cite{mark06qsgw,Kotani07,questaal_paper} (QS\emph{GW}), which, in contrast to conventional \emph{GW}, modifies
the charge density and is determined by a variational principle~\cite{Beigi17}, and finally an extension
QS$G\widehat{W}$ of QS\emph{GW}, where the polarizability needed to construct \emph{W} is computed including vertex
corrections (ladder diagrams) by solving a Bethe-Salpeter equation (BSE) for the two-particle
Hamiltonian~\cite{cunningham21}. The electron-hole two-particle correlations are incorporated within this
self-consistent ladder-BSE QS$G\widehat{W}$ implementation~\cite{Cunningham18,cunningham21} within the Tamm-Dancoff
approximation~\cite{tamm1,tamm2}.  As the present work is a study of excitons, we consider here only
QS$G\widehat{W}$ calculations.  A crucial difference in our implementation of
QS$G\widehat{W}$ from most other implementations of BSE is that the calculations are self-consistent in both self energy
$\Sigma$ and the charge density~\cite{swagcrx1,swagfirst}.  The effective interaction \emph{W} is calculated with ladder-BSE corrections and the
self energy, using a static vertex in the BSE.  \emph{G}, $\Sigma$ and \emph{W} are updated iteratively until all of
them converge.  Thus, in contrast to most recent \emph{GW}
studies~\cite{louie1,molina} our calculations are completely parameter free and have no starting
point bias.  In each cycle, the RPA polarizability is made anew, which
    determines the RPA \emph{W}.  In each cycle the four-point polarizability is recomputed from the (newly updated) static part of
    \emph{W}, to update $\widehat{W}$. Moreover, we checked the convergence in the QS$G\widehat{W}$ band gap~\cite{swagcrx1} and exciton
positions by increasing the size of the two-particle Hamiltonian. We increase the number of valence ($v$) and conduction
($c$) states and observe that for all materials the QS$G\widehat{W}$ band gap stops changing once 24 valence and 14
conduction states are included in the two-particle Hamiltonian. This stands in stark contrast to the
  weakly bound case. It reflects the atomic molecular-like picture where any state with significant orbital
  character participating in the exciton (Cr-3d or X-p) modify it, not just the band edge states as in the Wannier
  picture.  This allows us the flexibility to pin down the orbital characters that determine the position and
intensities of the excitons.

\begin{figure}
		\begin{center}
			\includegraphics[width=1\columnwidth, angle=-0]{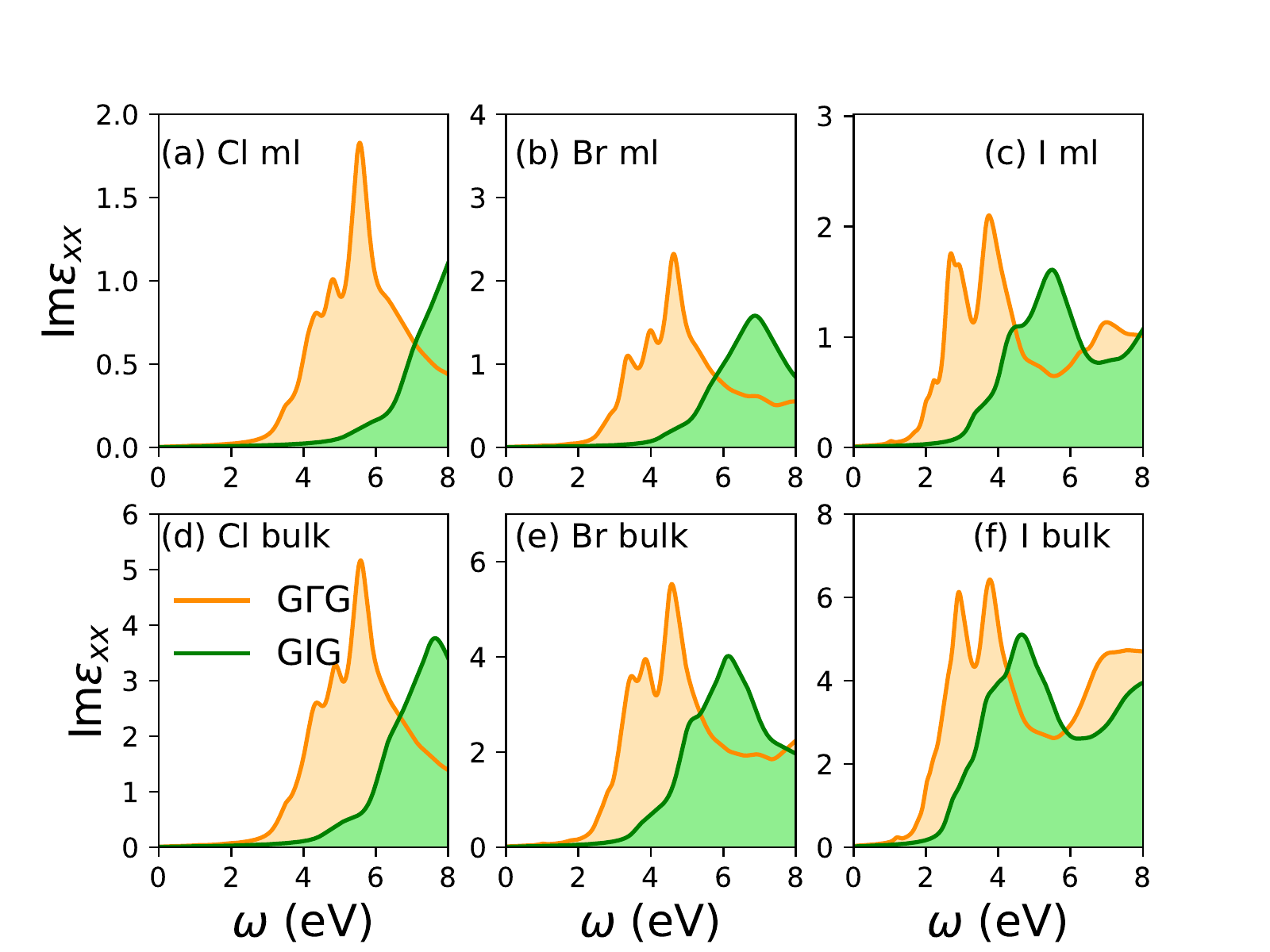}
			\caption{\ce{CrX3} : Imaginary part of macroscopic dielectric response Im$\epsilon_{xx}$ with
                          the perturbing electric field applied along the $(1 0 0)$ direction of the material, for free
                          standing monolayers (ml) of CrX$_{3}$ with (a) X = Cl, (b) X = Br , (c) X = I and bulk
                          CrX$_{3}$ with (d) X = Cl, (e) X = Br and (f) X = I.  The orange color stand for
                          QS$G\widehat{W}$ vertex $\Gamma$ corrected optical response, while the green color stands for
                          the vertex replaced by an identity matrix $I$ effectively making it a RPA optical response
                          with QS$G\widehat{W}$ one-particle eigenfunctions. Energy-dependent optical broadening that linearly varies from 2
                          milli-Hartree at $\omega=0$ and 50 milli-Hartree at $\omega=13.6$ eV is used.}
			\label{fig:excitonfull}
		\end{center}
\end{figure}

\begin{figure}
		\begin{center}
			\includegraphics[width=1\columnwidth, angle=-0]{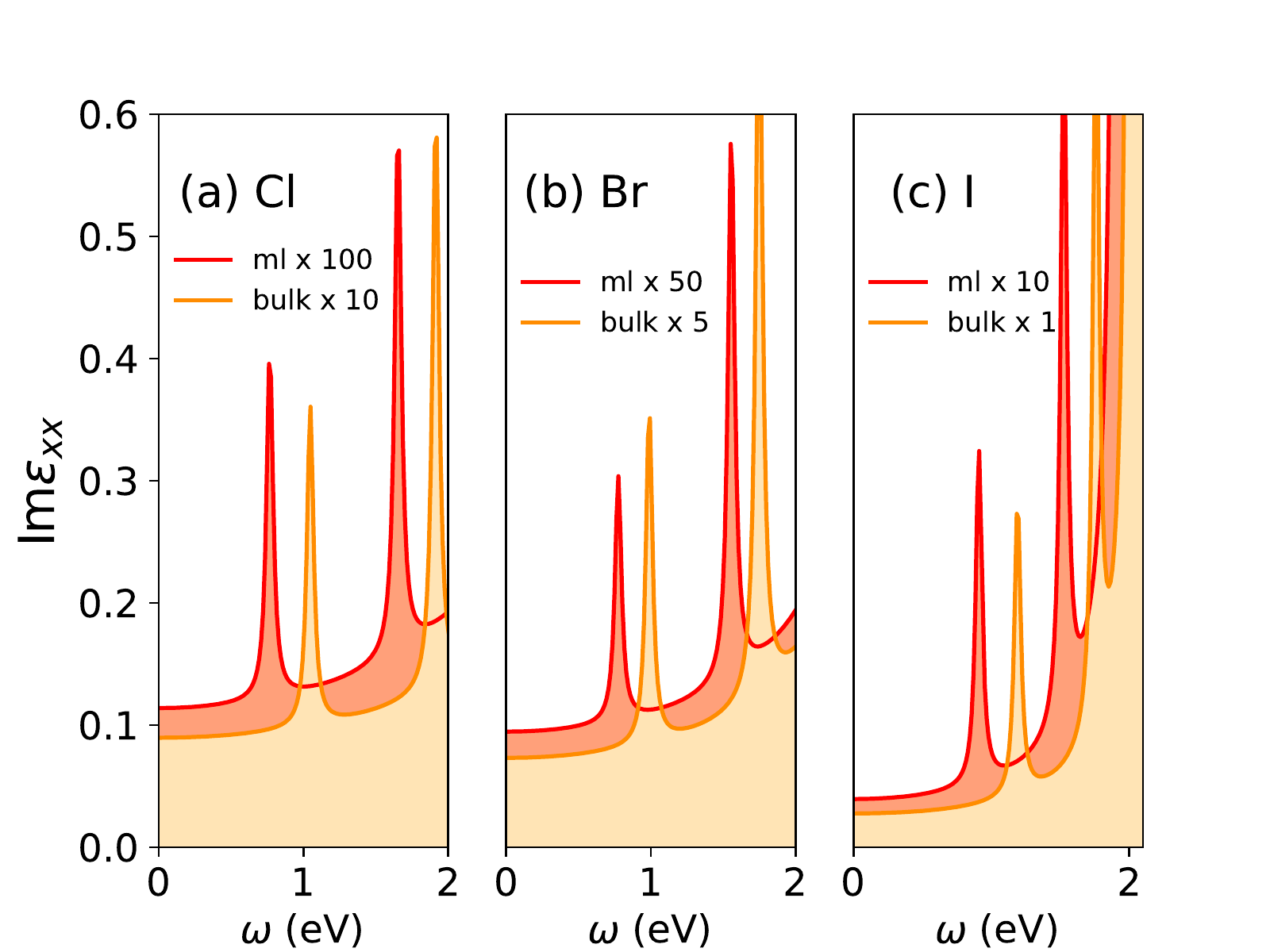}
			\caption{\ce{CrX3} : The low energy ($0-2$ eV) part of imaginary part of macroscopic dielectric response Im$\epsilon_{xx}$ with the perturbing electric field applied along the $(1 0 0)$ direction of the material,  for bulk and free standing monolayers of CrX$_{3}$ with (a) X = Cl, (b) X = Br , (c) X = I. Optical broadening of 2 milli-Hartree is used at all energies. The intensities of the peaks are multiplied by some constant factors to bring them to the same scale.}
			\label{fig:excitonzoom}
		\end{center}
\end{figure}

\section*{Results and Discussion}

Within QS$G\widehat{W}$, the Coulomb interaction $W$ softens in comparison to QS\emph{GW}, since the screening is
enhanced by the electron-hole attraction, as captured by the ladder diagrams.  The electronic band gap reduces roughly
by 20-25\% within QS$G\widehat{W}$ compared to QS\emph{GW} for all materials both in their bulk and monolayer
variants~(see Table \ref{tab:gaps}).  We compute the macroscopic dielectric response Im$\epsilon_{xx}$ with the
perturbing electric field applied along the $(100)$ direction of the material, with the QS$G\widehat{W}$ electronic
one-particle eigenfunctions assuming the optical vertex $\Gamma$=$I$ (identity matrix) and also using the explicitly
computed $\Gamma$. The excitons inside the one-particle band gap ($E \leq E_{g}$) can only be present in the second
case, and not in the first case where the optical transitions can only occur at or beyond the one-particle band gap ($E
\geq E_{g}$).

We observe that $\Gamma$ induces dramatic optical spectral weight transfer and series of excitons emerge inside the
one-particle band gap $\leq E_{g}$ in Fig.~\ref{fig:excitonfull}.  The optical weight transfers the most for X=Cl and
the least for X=I. A series of exciton peaks can be observed in all cases inside the gap. When we zoom into the low
energy ($0-2$ eV) part of the optical spectra we observe two deepest lying excitons ex$_{1}$ and ex$_{2}$ (see
Fig.~\ref{fig:excitonzoom}) that are robust across all cases and also for both bulk and monolayer variants of CrX$_{3}$.
We observe a weak redshift of $\sim$0.2 eV of these exciton peaks as the electronic screening reduces going from bulk to
monolayer. From the deepest lying exciton ex$_{1}$ position $E_{ex}$ and the electronic band gap $E_{g}$ at the
QS$G\widehat{W}$ level, we can compute the highest exciton binding energies as $E_{b} = E_{g} - E_{ex}$. E$_{b}$ enhances significantly in ml compared to the bulk, since lesser screening implies stronger exciton binding in ml. Note, although apparently E$_{ex}$ remains invariant in bulk and ml, both E$_{g}$ and E$_{b}$ enhances in ml compared to bulk. 

	\begin{table}[h]
		\footnotesize
		\begin{tabular}{c|@{\hskip 3pt}ccc@{\hskip 3pt}|@{\hskip 3pt}ccc}
			\hline
			theory & \multicolumn{3}{c}{Monolayer $E_{b}$ (eV)} & \multicolumn{3}{c}{Bulk $E_{b}$ (eV)} \cr
			& \ce{CrCl3} & \ce{CrBr3}  & \ce{CrI3} & \ce{CrCl3} & \ce{CrBr3} & \ce{CrI3} \cr
			\hline
			ex$_{1}$    & 4.75,2.62~\cite{molina}  & 3.85,2.05~\cite{molina},2.3~\cite{louie1}   & 1.95, 1.06~\cite{molina}   & 3.35  & 2.5 & 1.3  \cr
			ex$_{2}$  & 3.9  & 3.1   & 1.35   &  2.5  & 1.75 &  0.7 \cr
			ex$_{3}$   & 2.3  & 2.05   & 0.9    & 1.1 & 0.9 &  0.4 \cr
		\end{tabular}
		\caption{Exciton binding energies $E_{b}$ for three deepest lying excitons in all cases as computed from the difference between the one particle electronic band gap $E_{g}$ and position of the exciton peaks $E_{ex}$. We note the E$_{b}$ values as reported from LDA based single shot G$_{0}$W$_{0}$+BSE calculations for dielectric response performed by different groups.  In the supplemental material we discuss the possible reasons for their differences with our estimations.}
		\label{tab:binding}
	\end{table}

\begin{figure}
		\begin{center}
			\includegraphics[width=1\columnwidth, angle=-0]{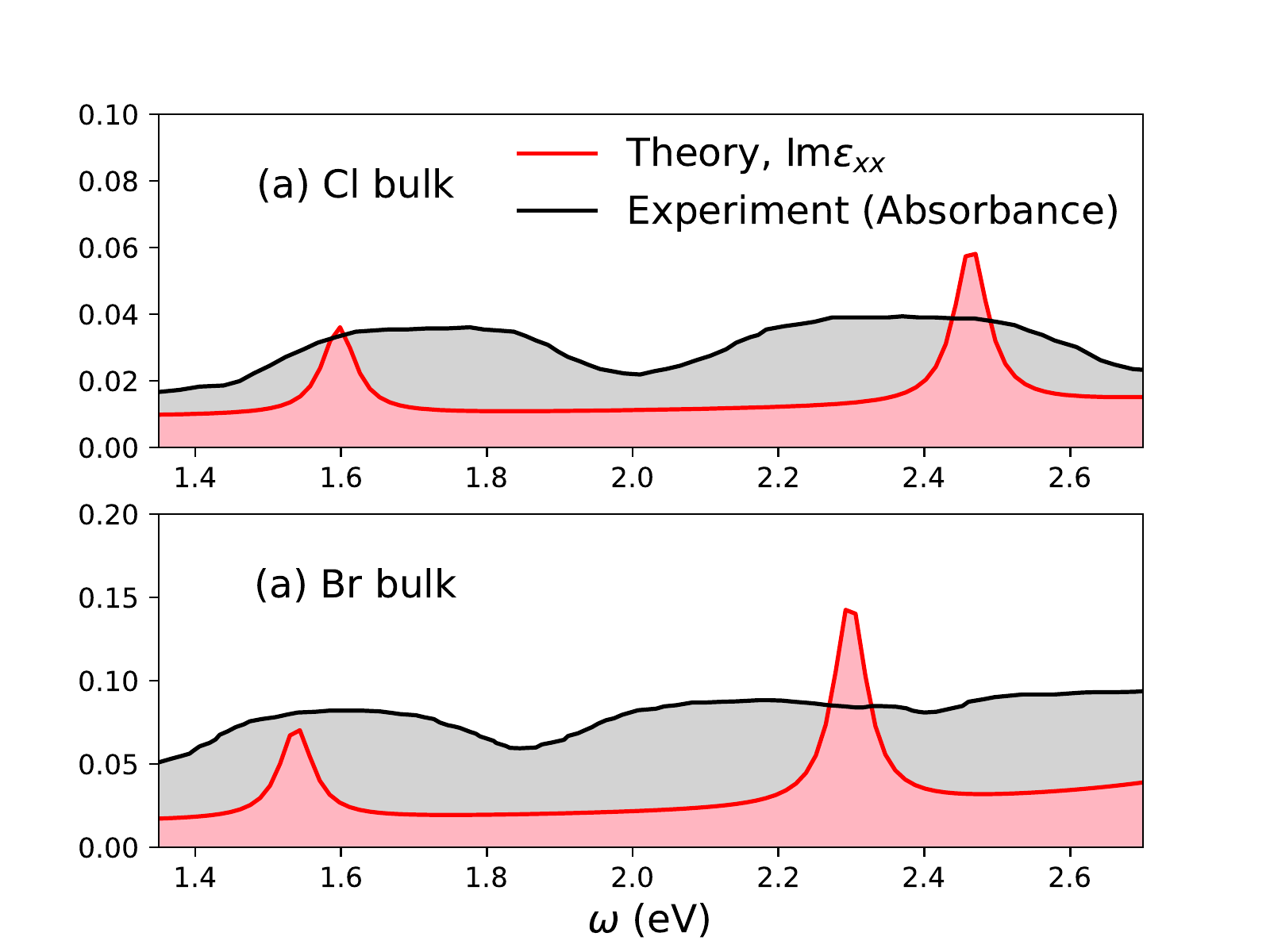}
			\caption{\ce{CrCl3,CrBr3} bulk :  Comparison between the adopted optical absorbance data from the recent experimental work on CrCl$_{3-x}$Br$_{x}$~\cite{abramchuk2018} against our theoretical results. The theoretical spectra for Im$\epsilon_{xx}$ is rigidly blue-shifted by $\sim$0.5 eV as our exciton peaks from $v$24$c$14 were too deep compared to experiments.  Note that need for such manual blue-shifting would not emerge required if we used the dielectric response computed from a smaller two-particle Hamiltonian, say $v$12$c$14 (see supplemental material). }
			\label{fig:exptclbr}
		\end{center}
\end{figure}

\begin{figure}
	\begin{center}
		\includegraphics[width=1\columnwidth, angle=-0]{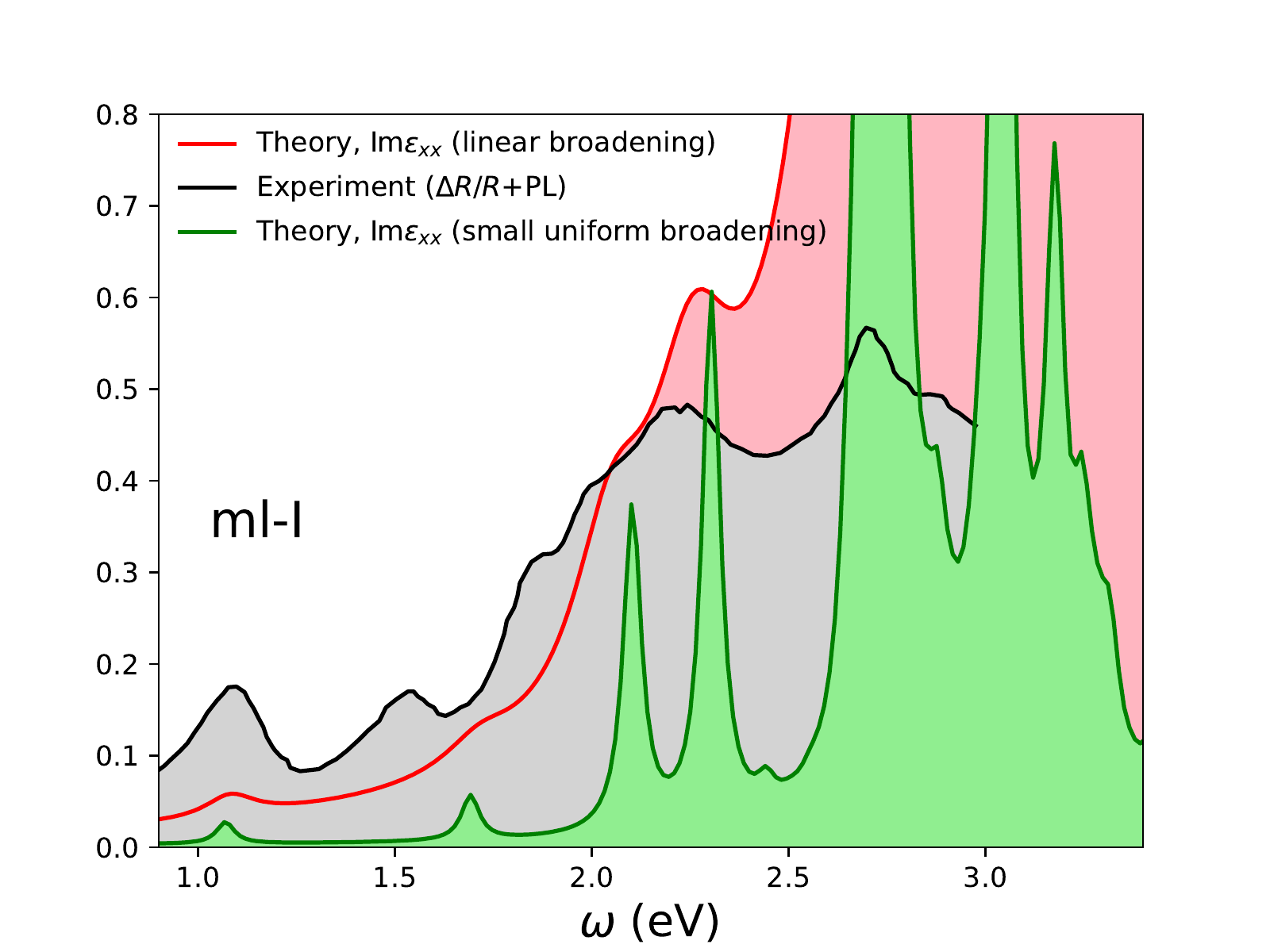}
		\caption{\ce{CrI3} monolayer : Comparison between the adopted differential reflectivity data combined with low energy photo-luminescence data from the recent experimental work on ml-CrI$_{3}$~\cite{seyler} against our theoretical Im$\epsilon_{xx}$ results. The theoretical spectra for Im$\epsilon_{xx}$ is rigidly blue-shifted slightly as our exciton peaks from $v$24$c$14 were by $\sim$0.2 eV too deep compared to experiments.  The theoretical data for different optical broadening schemes are also plotted on top of the adopted experimental data. }
		\label{fig:expti}
	\end{center}
\end{figure}

\begin{figure*}
	\begin{center}
\includegraphics[width=1\columnwidth, angle=-0]{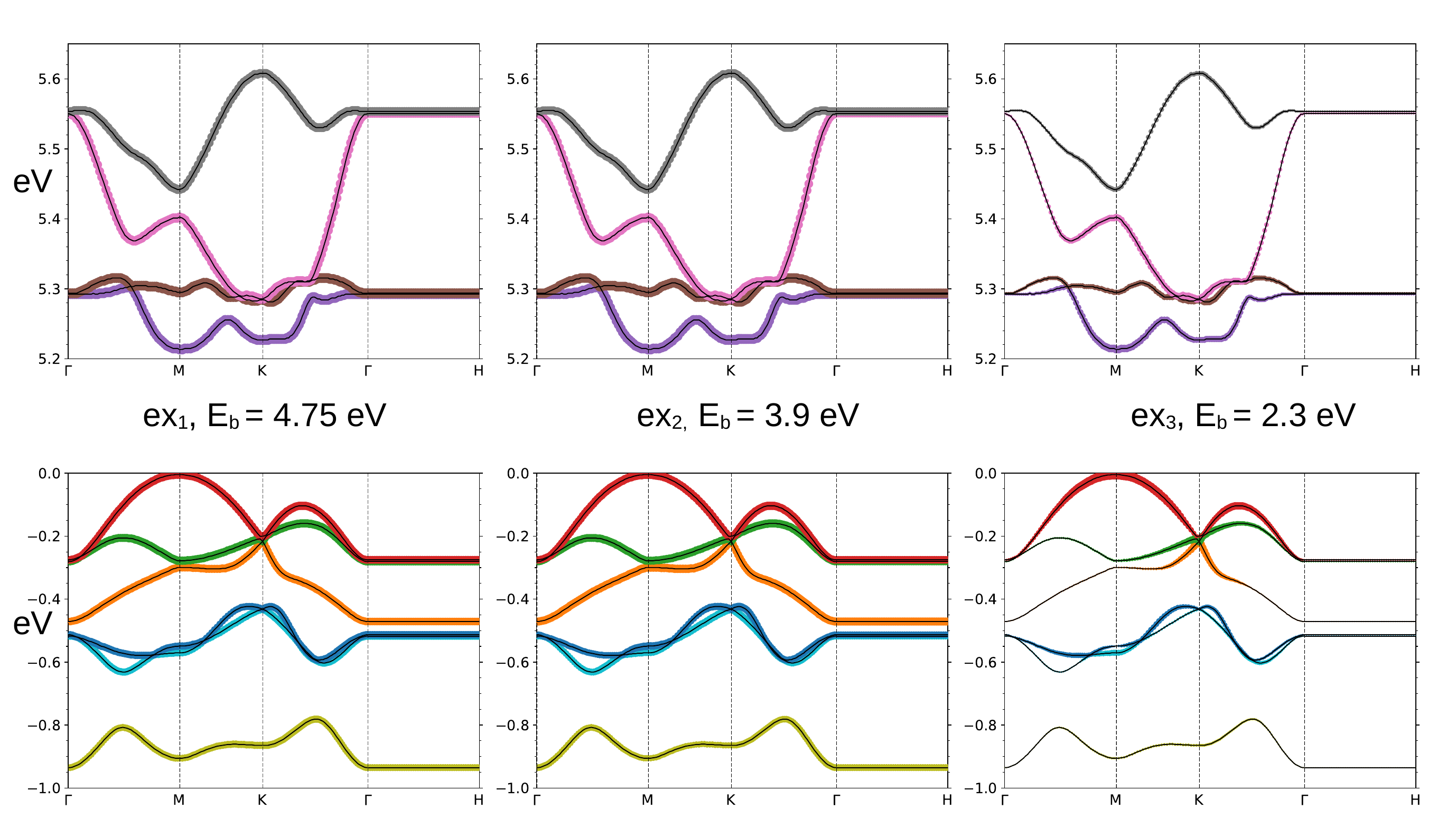}	
		\caption{\ce{CrCl3} Monolayer: The spectral weight analysis for the three deepest lying exciton ex$_{1}$, ex$_{2}$ and ex$_{3}$ from left to right respectively.  The size of the colored circles corresponds to the band contribution to the exciton spectral weight. Almost all the bands containing Cr-3d t$_{2g\uparrow}$ and e$_{g\uparrow}$ orbital characters contribute to ex$_{1,2}$. Exciton spectral weight is almost uniformly spread across these bands in case of ex$_{1,2}$. Spectral weight for ex$_{3}$, which has lesser binding energy compared to ex$_{1,2}$, becomes localized in band basis. Different colors are used to identify different bands. }
		\label{fig:excitoncl}
	\end{center}
\end{figure*}

In bulk, these two exciton peaks (ex$_{1,2}$) are experimentally well established in CrCl$_{3}$ and CrBr$_{3}$ from a
series of works~\cite{kamimura1966,grant1968,pollini1970,bermudez1979,nosenzo1984} performed between 1960's and 1980's.
In a recent work~\cite{abramchuk2018} on CrCl$_{3-x}$Br$_{x}$ the same peaks are observed again and also going from
$x$=3 to $x$=0 a weak blueshift in the peak positions can be noticed. Within our parameter free approach we can
reproduce the two-peaks in both $X$=Cl and Br and also their relative spacing and intensities agree almost perfectly
(see Fig.~\ref{fig:exptclbr}).  Some of the crucial features like weak blue shifting of the exciton peaks and change in
their relative spacing in CrCl$_{3}$ compared to CrBr$_{3}$ are also reproduced in our calculations. However, these two peaks in our calculations are $\sim$0.5 eV too deep compared to the experimental
findings. The position of these excitons change and they become systematically red-shifted as more screening channels
are included in the two-particle Hamiltonian we solve using QS$G\widehat{W}$. As we show in the Supplemental material the two peaks already form when the two-particle
Hamiltonian contains the minimal 6 valence bands ($v$) and 4 conduction bands ($c$), which are mostly of Cr-$d$
$t_{2g\uparrow}$ character and Cr-$d$ $e_{g\uparrow}$ character. So these exciton peaks are triplet in nature and emerge
from the 3$d$ $t_{2g-eg}$ transitions within the bands of mostly Cr character. They red-shift in tandem  as two-particle Hamiltonian includes more bands.


\begin{figure*}
	\begin{center}
\includegraphics[width=1\columnwidth, angle=-0]{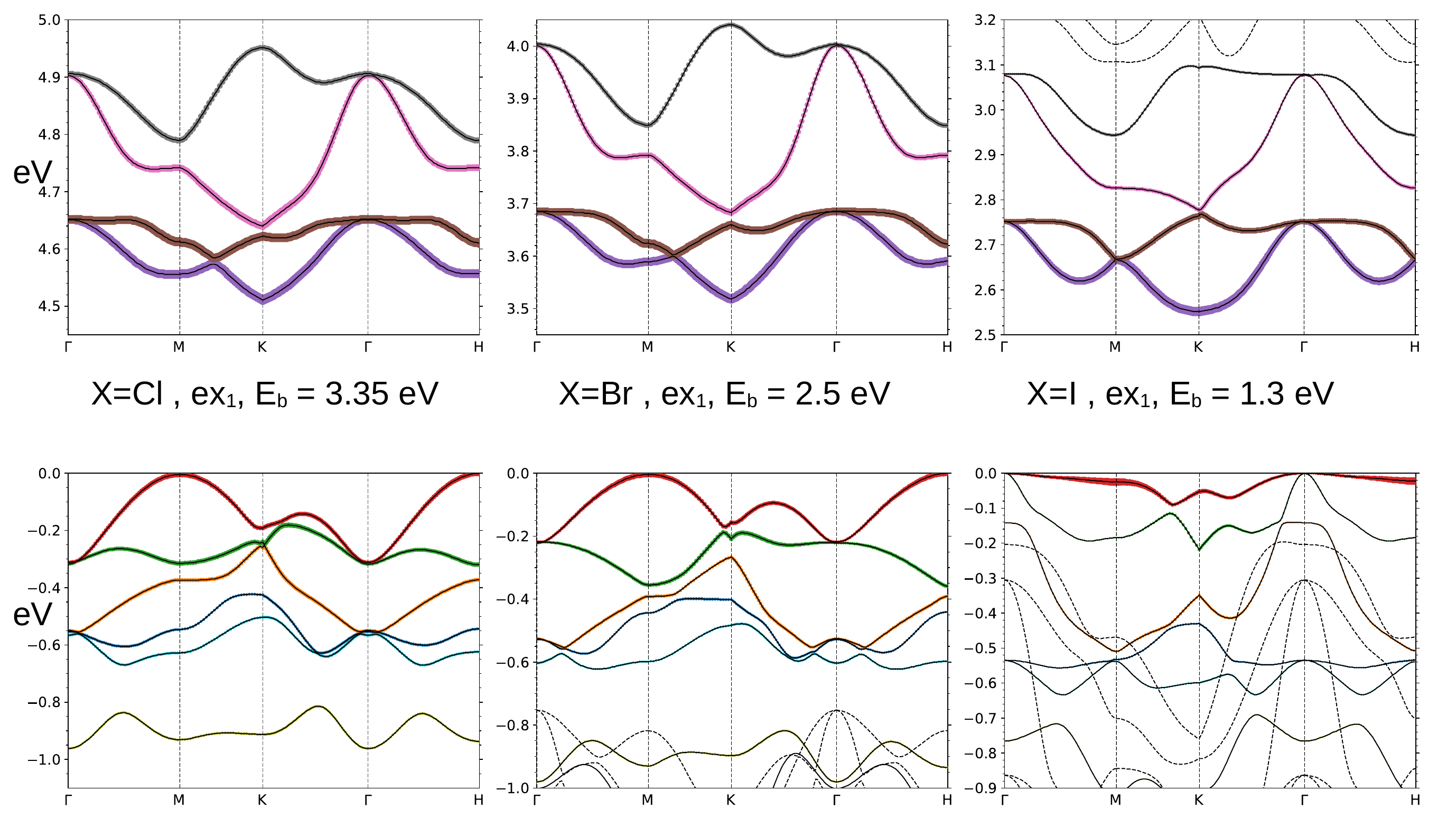}	
	\caption{\ce{CrCl3, CrBr3, CrI3} bulk : The spectral weight analysis for the deepest lying exciton ex$_{1}$ in each case. The spectral weight is raised to the fourth power to identify variations in the band contribution to the spectral weight. In the direction of Cl{\textrightarrow}Br{\textrightarrow}I the number of bands that contribute to the spectral weight of ex$_{1}$ decreases. Also the exciton spectral weight becomes more localized in band basis for X=I, compared to X=Cl,Br.}
		\label{fig:exciton}
	\end{center}
\end{figure*}

In a recent photoluminescence (PL) study on ml-CrBr$_{3}$ by Zhang \emph{et al.}~\cite{zhang}, the authors observe a peak at
1.35 eV.  They also observe that the PL peak energy is almost invariant of thickness of the sample (thickness ranging
between 6 and 73 nm), which suggests a localized transition.  Based on that they argue that this transition is
consistent with an atomic $d{-}d$ transition.  They further argue that that since the Laporte rule
prohibits such transition based on symmetry considerations, to relax the rule
symmetry breaking must be introduced via at least one mechanism, such as spin-orbit coupling, Jahn-Teller distortion and formation of odd-parity
phonons~\cite{seyler} as we noted in the introduction. They observe a broad PL linewidth and argue that it
serves as the evidence for the strong vibrionic coupling, resulting in photon sidebands. They also note that the
absorption peak that Dhillon \emph{et al.}~\cite{kamimura1966} observed at 1.67 eV in their work from 1966 in bulk CrBr$_{3}$ is
fundamentally the same peak Zhang \emph{et al.}~\cite{zhang} observes at 1.35 eV in ml-CrBr$_{3}$, which is assigned to the
absorption to the $^4T_{2}$ state at 1.5 K. They argue, further, that the Stokes shift of 320 meV, between absorption
and PL peaks of Dhillon \emph{et al.} and Zhang \emph{et al.} respectively, is due to the strong electron-lattice coupling. Similar
optical absorptions to that of Dillon \emph{et al.}~\cite{kamimura1966} is also confirmed in a recent study on bulk
CrBr$_{3}$~\cite{abramchuk2018}. 

Our theoretical results are fully consistent with experimental observations, without the need for
  spin-orbit coupling, lattice relaxations, or phonons.  We find that indeed, the low lying excitons ex$_{1}$ and ex$_{2}$ are robust in both bulk
  and monolayer, with modest differences in their position. 
We note in passing that the present simulations for
a free standing monolayer with a 60\,\AA\ vacuum, do not
  precisely correspond to a monolayer on a substrate.  The substrate effect is
  likely to be non-negligible in real systems, since even in the absence of significant covalent bonding
  to modify the energy band structure, the substrate will increase the dielectric response  which
  should reduce the excitonic binding energy. This could be one reason for the slight overestimate of our
    predicted binding energy in the ml case. However, later in the paper we discuss the intricate details of the vertex that we consider as one important reason for this overestimation in E$_{b}$. While the issue of the `exact position' of such excitons will be discussed in the following sections, clearly, their  intensity drops rapidly  from bulk to monolayer.  Note in Fig.~\ref{fig:excitonzoom} the monolayer intensity is scaled differently to compare against the bulk.  This is again fully consistent with experimental observations (see Fig.~1(b) in Zhang \emph{et
al}.~\cite{zhang}) and is probably one important reason why in ml-CrX$_{3}$ careful PL
experiments are needed to pick up these weak exciton intensities, while the conventional absorption or reflection
spectroscopy is enough to pick up the excitons in bulk. Nevertheless, our calculations establish that traditional explanations for the origin of these peaks are not needed,  but to
   understand the fundamental nature of these excitons a more careful analysis of the excitonic spectral weights and their
   origin are required.

Before proceeding, we attempt to benchmark our results for CrI$_{3}$ against the
existing body of literature. In a recent PL study on ml-CrI$_{3}$~\cite{seyler} an exciton peak is observed at 1.10 eV, and the position  hardly changes as the thickness of the sample is increased (see Fig.\,4(d) in Seyler \emph{et al.}~\cite{seyler}). From our calculations we also observe the
lowest energy exciton ex$_{1}$ position at 1.0 eV which is robust across bulk and monolayer variants, much like
CrCl$_{3}$ and CrBr$_{3}$.  We perform a thorough benchmarking of our theoretical optical spectra against what is
observed in recent experiments and the agreement is excellent for all energies up to
$\sim$3.5 eV (see Fig.~\ref{fig:expti}).  Also in the study by Seyler \emph{et al.} where the PL intensities are shown for the
1.10 eV peak for bulk, monolayer and layers of different thickness a dramatic drop in the PL intensity can be observed
as one goes from bulk to ML limit.  This is fully consistent with our calculations [see
Fig.~\ref{fig:excitonzoom}(c)]:  for the same material the ex$_{1,2}$ peaks can be $\sim$10
times more intense in bulk compared to its ML counterpart.  Among all these materials the excitons are the most intense
in CrI$_{3}$ (see Fig.~\ref{fig:excitonzoom}). This can be observed also in the old studies on the bulk
samples~\cite{kamimura1966,grant1968}.  This is again consistent with our calculations [see Figs.~\ref{fig:excitonzoom}(a)-\ref{fig:excitonzoom}(c)] where the exciton peaks are the
least intense in X=Cl and most intense in X=I. As we show, between ML-CrCl$_{3}$ and ML-CrI$_{3}$, 
the intensity of ex$_{1}$ varies by a factor of $\sim$10. This is a signature of the fact that although these two deepest lying excitons in all materials originate fundamentally
from transitions between Cr-$d$ $t_{2g\uparrow}$ and Cr-$d$ $e_{g\uparrow}$ states, their intensities are directly proportional to the hybridization of the atomic states with the environment. 
The (Cr-$d$, X-$p$) hybridization increases in the sequence Cl{\textrightarrow}Br{\textrightarrow}I, and is responsible
for the enhancement in the ex$_{1}$ intensity leading to brighter excitons.
X=Cl is the closest to the purely atomic scenario where the hybridization Cl-$p$--Cr-$d$ splitting is
  largest. To reach an unambiguous conclusion we need to perform further careful analysis of the exciton spectral
weights of all these excitons with
high-binding energies (ex$_{1,2,3}$) in the entire class of these bulk and ML magnets. This allows us the opportunity to
pin down both their fundamental nature and sensitivity to the ligand states.

We analyze which band pairs and k-points contribute to the Im$\epsilon_{xx}(\omega)$ exciton spectrum integrated
over narrow energy ranges near the peaks of the exciton spectrum. We divide the QS$G\widehat{W}$ exciton region in
separate intervals depending on where the three deepest exciton peaks reside. For example in ML-CrCl$_{3}$ the windows
would be $[0.6-1.4]$, $[1.45-2.25]$, and $[2.5-3.36]$ eV, corresponding each to a separate peak in the optical
spectrum. Fig.~\ref{fig:excitoncl} shows which band states $(n, k)$ contribute to the exciton eigenvalues.  We obtain
this figure by using the eigenvectors $|A^{\lambda}_{n,k}|^{2}$ of the two-particle Hamiltonian as a weight at each $k$ and band $n$ where $\lambda$
indicates the exciton eigenvalue and then including all eigenvalues in a given energy range. This is then visualized as
the size of the circles on the band structure. The valence bands of primarily of
Cr-$d$ $t_{2g\uparrow}$ characters and conduction bands of primarily Cr-$d$ $e_{g\uparrow}$ characters contribute to the
ex$_{1}$ formation in the window $[0.6-1.4]$ eV. Intriguingly enough, 10 entire bands ($v$6 and $c$4) contribute almost
uniformly (see Fig.~\ref{fig:excitoncl}) across the Brillouin zone to the ex$_{1}$ spectral weight. This is true for
ex$_{2}$ in the window of $[1.45-2.25]$ as well, except for parts of the conduction and valence bands (third and forth
conduction bands counted from the conduction edge, third, forth, fifth and sixth valence bands counted from the valence
band top) contribute slightly less compared to other bands.  Thus, bands contributing the least to
  the spectral weight have the most Cl-$p$ character.    However,  the same
analysis for ex$_{3}$ shows that mostly the top most valence band and bottom most
conduction band contribute to the exciton spectral weight, while the other bands weakly participate in the process. This
gives clear indication that in CrCl$_{3}$ ex$_{1}$ and ex$_{2}$ are significantly localized in real space and are of
Frenkel nature, while ex$_{3}$ is closer to the Wannier-Mott limit. We also note the situation remains
invariant whether we perform similar analysis for bulk or ML cases.

\begin{figure}
	\begin{center}
\includegraphics[width=1.0\columnwidth, angle=-0]{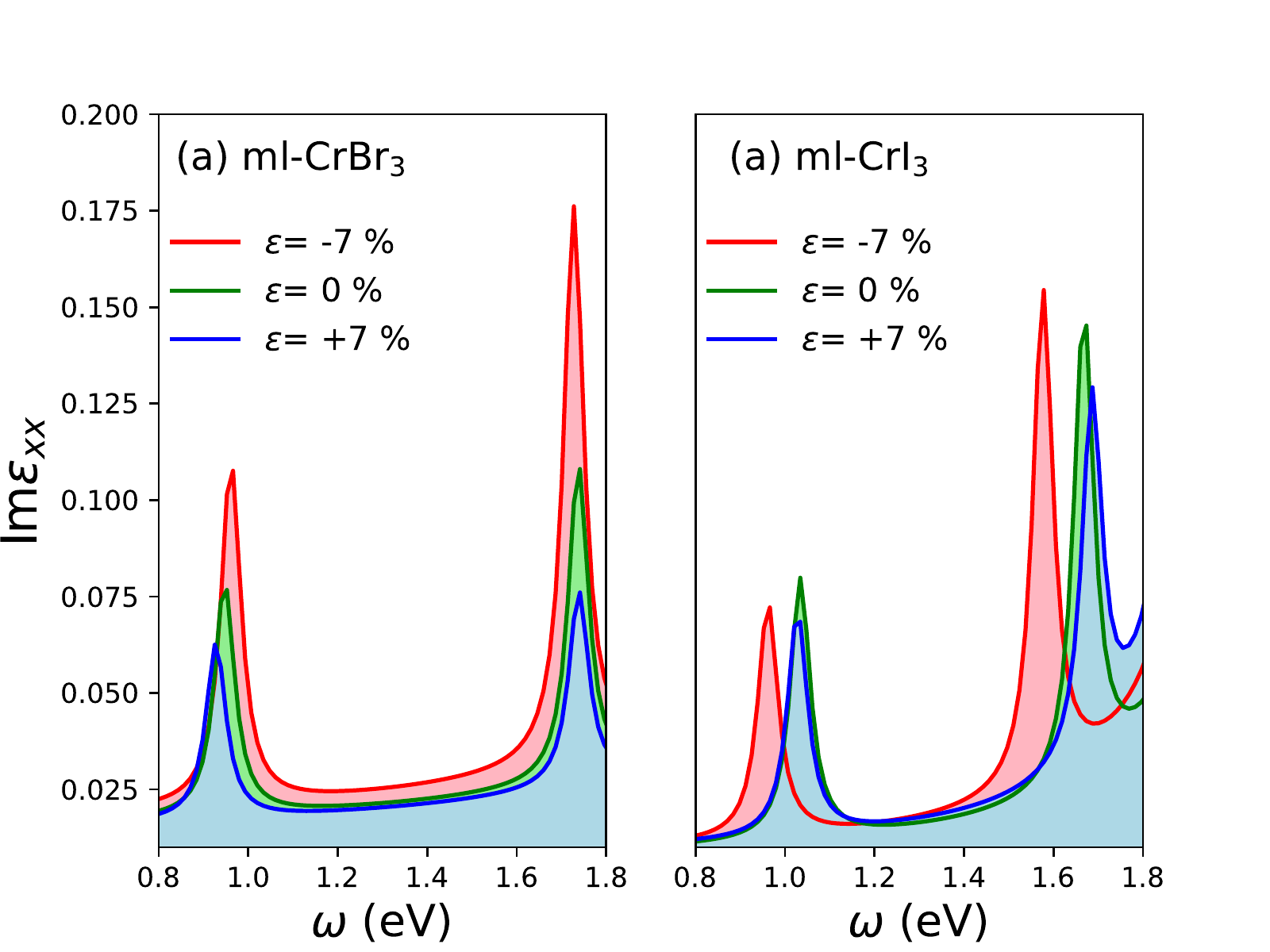}	
	\caption{\ce{CrBr3, CrI3} Monolayer : Volume conserving tensile and compressive strains $\epsilon$ are applied. $E_{ex}$ does not change for X=Br , while it changes by 100 meV for compressive strain of $\epsilon$= 7\% in X=I. Intensities of the peaks are multiplied by a constant factor of three in ml-CrBr$_{3}$ to bring them to the same scale as ml-CrI$_{3}$.}
		\label{fig:strain}
	\end{center}
\end{figure}

Now, we analyze the same across the series Cl{\textrightarrow}Br{\textrightarrow}I. We raise the  spectral weight to the fourth power in all cases to identify variations in the band contribution to the spectral weights. We observe that for ex$_{1}$ in ML-CrCl$_{3}$, while almost all of 6 valence bands and 4 conduction bands contribute uniformly across the BZ, the situation is quite different for the ex$_{1}$ in CrI$_{3}$. In case of ML-CrI$_{3}$, we observe that mostly the two bottom most conduction bands and one top most valence band contribute to ex$_{1}$ spectral weight (see Fig.~\ref{fig:exciton}), suggesting that the ex$_{1}$ in CrI$_{3}$ is significantly delocalized in nature compared to the other extreme of CrCl$_{3}$. Also, even for the top most valence band, we can identify the red circles becoming larger closer to the M-point and fainter away from that. A careful analysis of the orbital component of this top most valence band shows~\cite{swagcrx1} that the Cr-$d$ orbital character is most prominent at and around M-point while it becomes more I-$p$ like away from it. Altogether, CrI$_{3}$ emerges as the extreme case of this series where ex$_{1}$ is partially localized in momentum space, in strong contrast to CrCl$_{3}$. Similar analysis for both bulk and ML for all cases are performed and the results are shown in the supplemental materials. 

\begin{figure}
	\begin{center}
		\includegraphics[width=1\columnwidth, angle=-0]{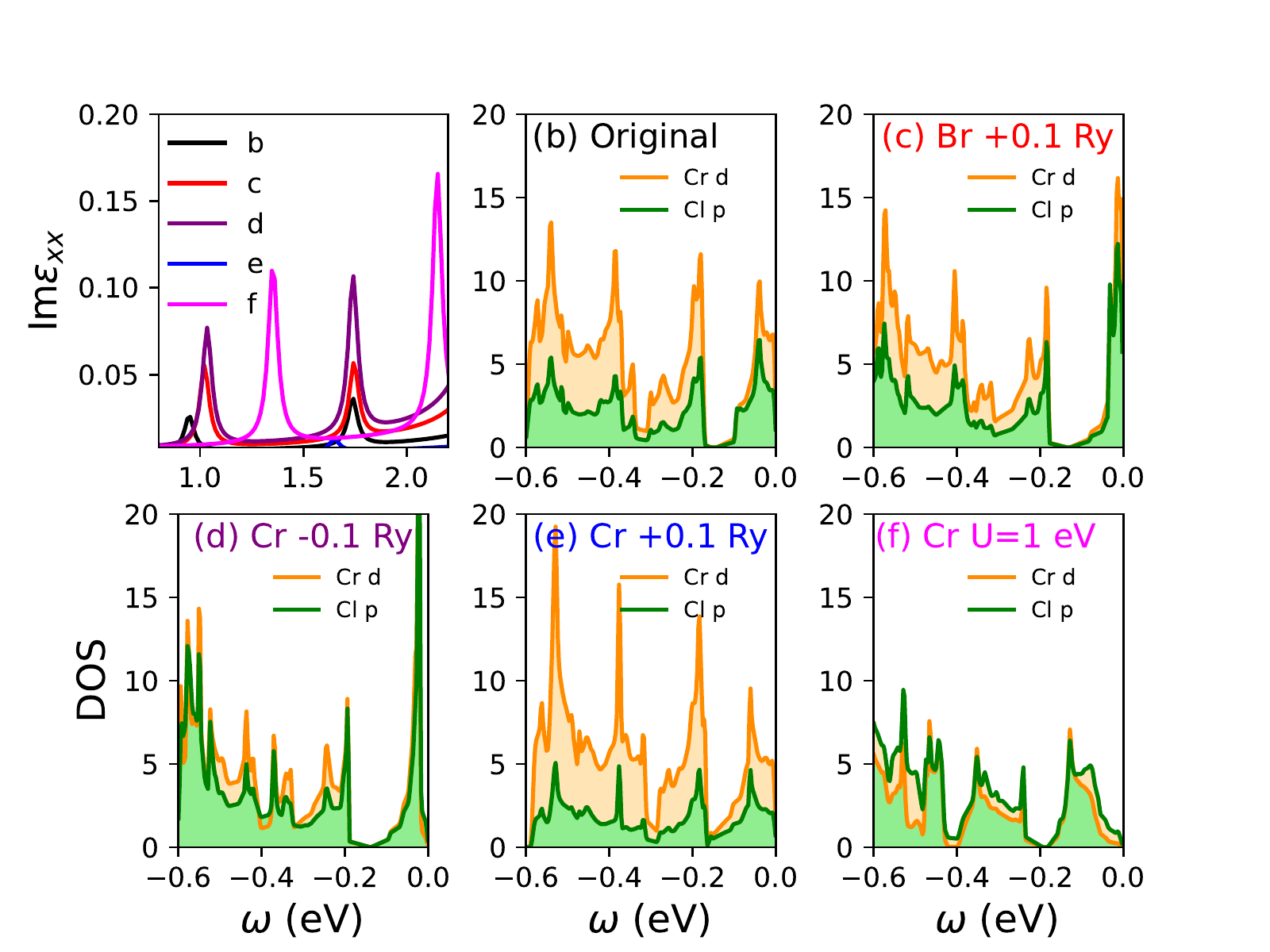}
		\caption{\ce{CrBr3} Monolayer: QS$G\widehat{W}$ results in different circumstances.  $Br +0.1 Ry$ shifts the Br-p centre of mass up by 0.1 Ry, $Cr -0.1 Ry$ and $Cr +0.1 Ry$ down/up shifts Cr-d the centre of mass by 0.1 Ry and $Cr U=1 eV$ adds U=1 eV Cr-d. In cases with larger Cr-d and Br-p hybridization the excitons become more intense and get blue shifted and the reverse happens for lesser hybridization.  However, the position of ex$_{1,2}$ changes dramatically for $Cr U=1 eV$.}
		\label{fig:position}
	\end{center}
\end{figure}

To further explore the crucial role of Cr-$d$-X-$p$ hybridization in determining the intensities and positions of these excitons, we apply a range of weak perturbations to the QS$G\widehat{W}$ two-particle Hamiltonian that modifies the hybridization. Such a physical situation could be simulated within our calculations by either down/up-shifting the X-p centre of mass, down-shifting/up-shifting the Cr-d centre of mass or by adding U to the Cr-d that shifts the Cr majority spin sector down and the minority spin sector up. We try these options on ml-CrBr$_{3}$: shift the centre of mass of Br-p up by 0.1 Ry, shift the Cr-d centre of mass down/up by 0.1 Ry and add U=1 eV to Cr-d. We refer to them respectively as $Br +0.1 Ry$, $Cr -0.1 Ry$, $Cr +0.1 Ry$  and $Cr, U=1 eV$ for the rest of the discussion. The band gap changes in all cases, so does the hybridization between Cr-d and X-p. For  $Cr +0.1 Ry$ (see Fig.~\ref{fig:position}(e)) the hybridization reduces slightly compared to the original unperturbed case (see Fig.~\ref{fig:position}(b)) and that leads to less intense exciton peaks (see the blue and black curves in  Fig.~\ref{fig:position}(a)). Also as the atomic nature of the Cr-$d$ becomes more prominent and it reflects in the exciton peaks getting weakly red shifted (larger exciton binding energy E$_{b}$) compared to the unperturbed scenario.  The reverse happens with $Br +0.1$ and for $Cr -0.1 Ry$. In both the cases the intensity of the ex$_{1,2}$ peaks increase significantly (see Fig.~\ref{fig:position}(a) red and purple curves) which is a direct consequence of larger hybridization between Cr-d and Br-p (see Fig.~\ref{fig:position}(c,d)) while the peak positions also get weakly blue shifted leading to lesser E$_{b}$. However, with $Cr, U=1 eV$ the ex$_{1}$ gets significantly blue shifted to 1.35 eV (see Fig.~\ref{fig:position}(a) magenta curve). Note that 1.35 eV is the exact position of the ex$_{1}$ from the recent PL study~\cite{zhang} on ml-CrBr$_{3}$).  In this scenario the valence band manifold becomes dominated by the Br-$p$ (see Fig.~\ref{fig:position}(f)). This also suggests that the missing details of the $\Gamma$, whether it is the dynamics in $\Gamma$ or the missing $\Gamma$ itself from the $\Sigma$, will most likely lead to the same qualitative effect leading to the correction in the absolute positions of these deep lying excitons. Within our QS$G\widehat{W}$ framework, vertex is explicitly included in $W$ and is absent from $\Sigma$. One primary effect of including $\Gamma$ in $\Sigma$ is to modify the relative centre of masses of Cr-d and X-p. Note that our E$_{ex_{1}}$ estimation was $\sim$ 0.5 eV off (too deep) in both bulk and ml. With  $Cr, U=1 eV$ we can correct E$_{ex_{1}}$ by $\sim$0.5 eV in both bulk and ml. Hence, the discrepancy between our theory and experiment is related to the exact nature of the vertex and not the dimensionality of the material. This is also intuitive that this discrepancy is large ($\sim$0.5 eV) in X=Cl,Br where the systems are most atomic in nature and smaller ($\sim$0.2 eV) in X=I, where it is least atomic or most band-like. It is expected that the dynamics in vertex or a better vertex altogether (possibly from approaches that are more `exact' than these many body perturbative approaches) is more relevant in the atomic scenario.

To further verify the Frenkel-ness of the excitons, we apply moderate volume conserving strain $\epsilon$ on the monolayers of CrBr$_{3}$ and CrI$_{3}$. We observe that the positions of the excitons ex$_{1,2}$, $E_{ex}$ remain invariant with both tensile and compressive strains in X=Br while it changes in X=I by roughly 100 meV for compressive strain $\epsilon$ = 7\%. Weak changes with tensile $\epsilon$ in X=I can also be observed. This is consistent with our conclusions of excitons being most Frenkel like in X=Cl,Br and more Wannier-Mott like in X=I.  


While what we should characterize as a proper Frenkel exciton is a matter of definition, it becomes apparent from our
analysis that even for the deepest lying exciton with the largest binding energy, it is most Frenkel like in CrCl$_{3}$
and the least Frenkel like in CrI$_{3}$. We show that the lighter the halogen is, the weaker is the intensity of these
Frenkel excitons making it even more challenging experimentally to observe them.  Nevertheless, for a given material, as
we look for excitons with lesser binding energies, for example ex$_{3}$, in all cases, is the closest analogue of
Wannier-Mott exciton, in the sense of what is observed in, say, LiF or MoS$_{2}$. The degree of `Wannier-Mott'-ness is
proportional to the degree of hybridization between the Cr-$d$ and X-$p$ states, thereby supporting  \emph{the picture}
of a `band-origin' for these excitons in CrX$_{3}$. The band-origin also suggests that it is, probably, possible to tweak
their binding energies and intensities by applying shear, strain, magnetic field or simply by scanning across the
periodic table looking for elements with varying degree of hybridization. However, as we show tensile or compressive strains lead to no (in X=Cl,Br) or weak changes (in X=I) to the exciton positions, suggesting that effectively these deep lying excitons across this entire class of systems are mostly Frenkel like. One interesting aspect would be to
realize crystals made of Cr and F, and/or Cr and O in the same crystalline space-group as the rest.

\section*{Conclusions}

Our work summarily excludes the stringent requirement of relaxation of the {Laporte Rule}, that forbids the `atomic'
$d{-}d$ transitions based on symmetry arguments, as a fundamental principle of origin of these excitons in CrX$_{3}$.
While the deepest lying excitons are mostly localized on the Cr-d orbitals, they delocalize within the
unit cell on the X-$p$ orbitals as well. The degree of delocalization and real-space range of these excitons increase as
the halogen atoms become heavier and contains more core states that are shallower compared to the Cr-$d$ states. This
suggests, if CrF$_{3}$ exists that could host the most localized Frenkel-like excitons, with the largest binding
energies, from the entire series with, probably, the closest analogue of what can be characterized as purely atomic
$d{-}d$ transition.  Nevertheless, in real world and crystalline environment there should always exist finite, albeit
small, hybridization between different angular momentum states, leading to invariably `band-origin' of excitons in all
cases. We explicitly show that it is possible to modify both the intensities and positions of these excitons by applying weak perturbation to the Hamiltonian that changes the hybridization environment.

\section*{Acknowledgements} 
MIK, ANR and SA are supported by the ERC Synergy Grant, project 854843 FASTCORR (Ultrafast dynamics of correlated
electrons in solids). This work was authored (in part) by the National Renewable Energy Laboratory, operated by Alliance for Sustainable Energy, LLC, for the U.S. Department of Energy (DOE) under Contract No. DE-AC36-08GO28308. Funding was provided by the Theoretical Condensed Matter Physics program within the Office of Basic Energy Sciences. The views expressed in the article do not necessarily represent the views of the DOE or the U.S. Government. The U.S. Government retains and the publisher, by accepting the article for publication, acknowledges that the U.S. Government retains a nonexclusive, paid-up, irrevocable, worldwide license to publish or reproduce the published form of this work, or allow others to do so, for U.S. Government purposes.
We acknowledge PRACE
for awarding us access to Irene-Rome hosted by TGCC, France and Juwels Booster and Cluster, Germany.

\section*{method}
Single particle calculations (DFT, and energy band calculations with the static quasiparticlized QS\emph{GW} self-energy
	$\Sigma^{0}(k)$) were performed on a 12$\times$12$\times$1  \emph{k}-mesh while the (relatively smooth) dynamical self-energy
	$\Sigma(k)$ was constructed using a 6$\times$6$\times$1 \emph{k}-mesh and $\Sigma^{0}$(k) extracted from it.  For each
	iteration in the QS\emph{GW} self-consistency cycle, the charge density was made self-consistent.  The QS\emph{GW} cycle
	was iterated until the RMS change in $\Sigma^{0}$ reached 10$^{-5}$\,Ry.  Thus the calculation was self-consistent in
	both $\Sigma^{0}(k)$ and the density.  Numerous checks were made to verify that the self-consistent $\Sigma^{0}(k)$ was
	independent of starting point, for both QS$GW$ and QS$G\widehat{W}$ calculations; e.g. using LDA or Hartee-Fock self-energy as the initial self energy for QS\emph{GW} and using LDA or QS\emph{GW} as the initial
	self-energy	for QS$G\widehat{W}$. We achieve these optical spectra in Fig.~\ref{fig:excitonzoom} using a uniform optical
broadening of 2 milli-Hartree at all energies, so that the peaks can be identified clearly. By
contrast, in Fig.~\ref{fig:excitonfull} an energy-dependent broadening that linearly varies from 2 milli-Hartree at the lowest energies to maximum of 50
milli-Hartree at the highest energy (at 1 Rydberg) was used.

\section*{Competing interests}
The authors declare no competing financial or non-financial interests.
\section*{Correspondence}
All correspondence, code and data requests should be made to SA.
\section*{Data Availability}

All input/output data can be made available on reasonable request. All the input file structures and the commandlines to launch calculations are rigorously explained in the tutorials available on the Questaal webpage~\cite{questaal_web} \href{https://www.questaal.org/get/}.

\section*{Code Availability}
The source codes for LDA, QS\emph{GW} and QS$G\widehat{W}$ are available from~\cite{questaal_web}  \href{https://www.questaal.org/get/}  under the terms of the AGPLv3 license.

\section*{Supplemental Material}

In the supplemental material we benchmark our results with previous theoretical and experimental results, perform the spectral weight analysis for the excitons in bulk and monolayer variants and convergence in $E_{ex}$ with varying two-particle Hamiltonian sizes.  

\subsection*{Benchmarking against existing theoretical and experimental results:}
The recent
work by Wu \emph{et al}.~\cite{louie1} predicts $E_{b}$=2.3 eV in monolayer CrBr$_{3}$ for the deepest lying bright
exciton. From their one shot L(S)DA+\emph{GW} calculations they achieve $E_{g}=3.8$ eV.  However, when they solve the
\emph{GW}+BSE to find the deepest lying bright exciton ex$_{1}$, the peak is observed at $E_{ex}=1.5$ eV, thereby
leading to $E_{b}=2.3$ eV. Note that our estimation of $E_{b}=2.5$ eV from bulk-CrBr$_{3}$ is very similar to their
estimation, nevertheless, our estimation of $E_{b}=3.85$ eV from ml-CrBr$_{3}$ is much higher. The primary reason for
this originates from the difference in the one-particle gap (4.65 eV in QS$G\widehat{W}$ compared to 3.8 eV
  from DFT+$U$+\emph{GW}). Also, the two-particle Hamiltonian that we solve within QS$G\widehat{W}$ contains 24 valence
and 14 conduction states, leading to $E_{ex}=0.8$ eV.  As we show below, a modestly different $E_{ex}$ is
  obtained when we use 21($v$)+14($c$) as Wu \emph{et al.} did.  The absolute position of $E_{ex}$ is
  much slower to converge than the relative splittings of the two bright excitons, as we show below.  This can be simply understood in terms of the orbitals contributing to the excitons: the highest valence bands contain the 6 Cr-3$d$-$t_{2g\uparrow}$ states and 18 X-$p$ states (the unit
cell contains 2 Cr atoms and 6 X atoms). These hybridize and all of them contribute to the exciton. Wu \emph{et al}.~\cite{louie1} argues that unlike monolayer transition metal
dichalcogenides where the lowest-energy bright excitons are of Wannier type with a diameter of several
nanometers~\cite{wannierexciton,wannierexciton1}, ML-CrBr$_{3}$ hosts bright charge-transfer exciton states that extend over a few
primitive cells, indicating formation of band transitions instead of intra-atomic $d{-}d$
transitions. They further argue that exciton distribution in ML-CrBr$_{3}$ is consistent with the intuition that a larger exciton binding energy is associated with a smaller exciton radius~\cite{cohen2016}. Their
findings for ML-CrBr$_{3}$ are consistent with our present observations, and indeed they extend to the entire CrX$_{3}$ (X=Cl,Br,I) family, both ML and bulk.

Further, we benchmark our results against the study by Molina \emph{et
al.}~\cite{molina} who uses a very similar method like Wu \emph{et al.}~\cite{louie1}, namely, L(S)DA+\emph{GW}. Also in both the
works~\cite{louie1,molina} the choice of the L(S)DA eigenfunctions as starting eigenfunctions for their single-shot
\emph{GW} calculations is completely consistent as both uses $U=1.5$ eV and $J=0.5$ eV. Molina \emph{et al.}~\cite{molina}
achieves $E_{g}=4.45$ eV for ml-CrBr$_{3}$ and $E_{ex}=2.40$ eV, leading to $E_{b}=2.05$ eV.  The fact that the
$E_{ex}$, in this study by Molina \emph{et al.}, is significantly blue shifted ($\sim$1 eV compared to Wu \emph{et al.}~\cite{louie1}
and $\sim$1.6 eV compared to our work) could be due to many reasons, the primary of which we think is the incomplete
convergence in the two-particle Hamiltonian size. However, we could not find any comment in these two works about the
convergence check in terms of the vacuum size. As we showed in our previous work~\cite{swagcrx1} the vacuum correction,
further, can lead to enhancement in $E_{g}$ in ml-CrBr$_{3}$ by $\sim$0.4 eV, and hence, could be a natural explanation
for an enhanced estimation of $E_{b}$ in our work.

\begin{figure}
		\begin{center}
			\includegraphics[width=1\columnwidth, angle=-0]{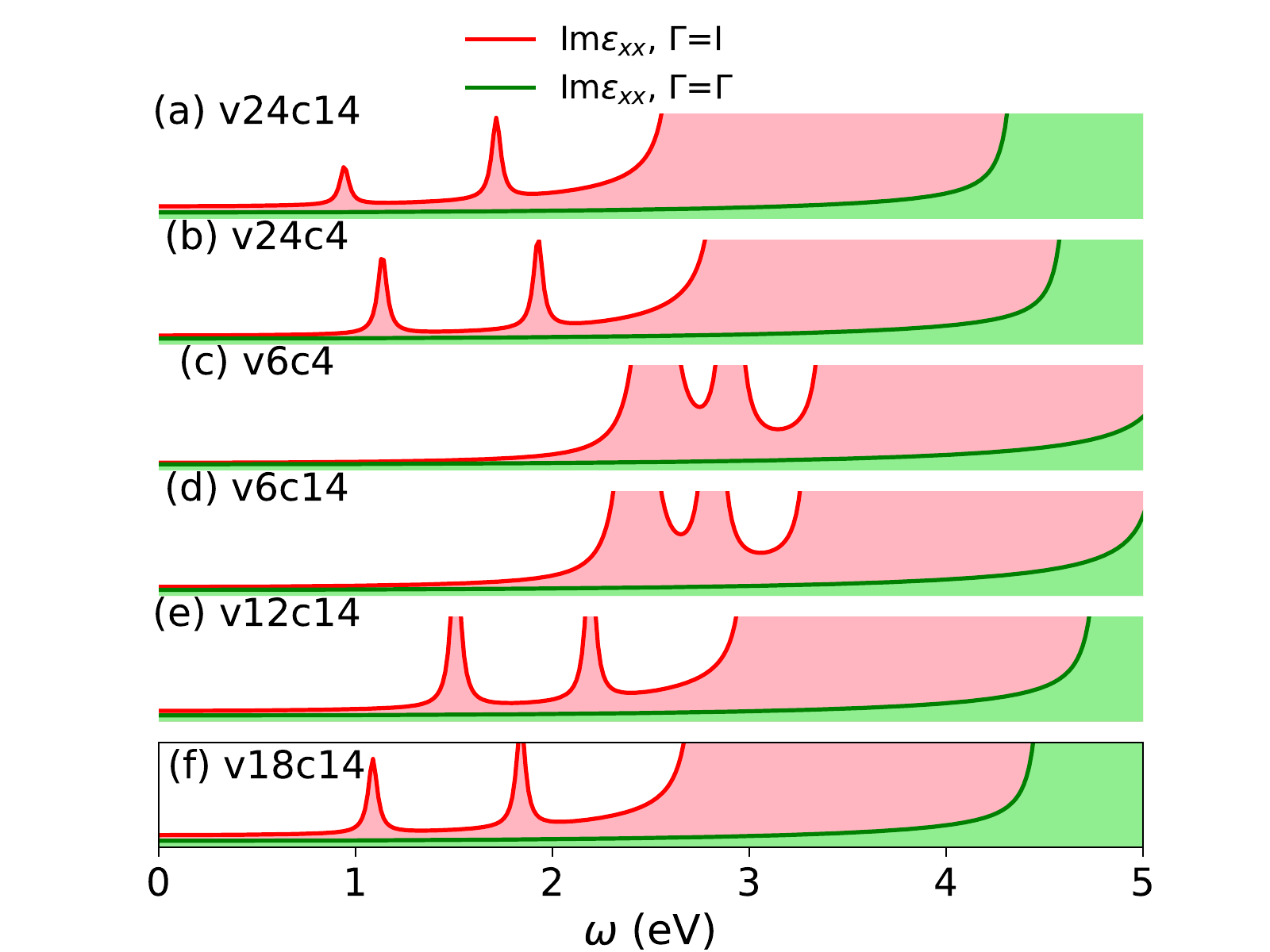}
			\caption{\ce{CrBr3} Monolayer:  Covergence in $E_{ex}$ with the two-particle Hamiltonian size by varying the numbers of valence ($v$) and conduction bands ($c$). The green curves that are for the RPA-Im$\epsilon_{xx}$ (the vertex $\Gamma$ put to $I$) also changes for different $v$-$c$ choices, since the one-particle eigenfunctions change in each case. The exciton peaks and the optical spectral weight gets redshifted with larger $v$-$c$ choices, as more screening channels are included. The $E_{ex}$ is most sensitive to $v$ and weakly depends on $c$ for $c$ $\geq$4 (the critical number of bands that contain mostly the Cr-$d$-$e_{g\uparrow}$ states).}
			\label{fig:hamilbr}
		\end{center}
\end{figure}

\begin{figure}
		\begin{center}
			\includegraphics[width=1\columnwidth, angle=-0]{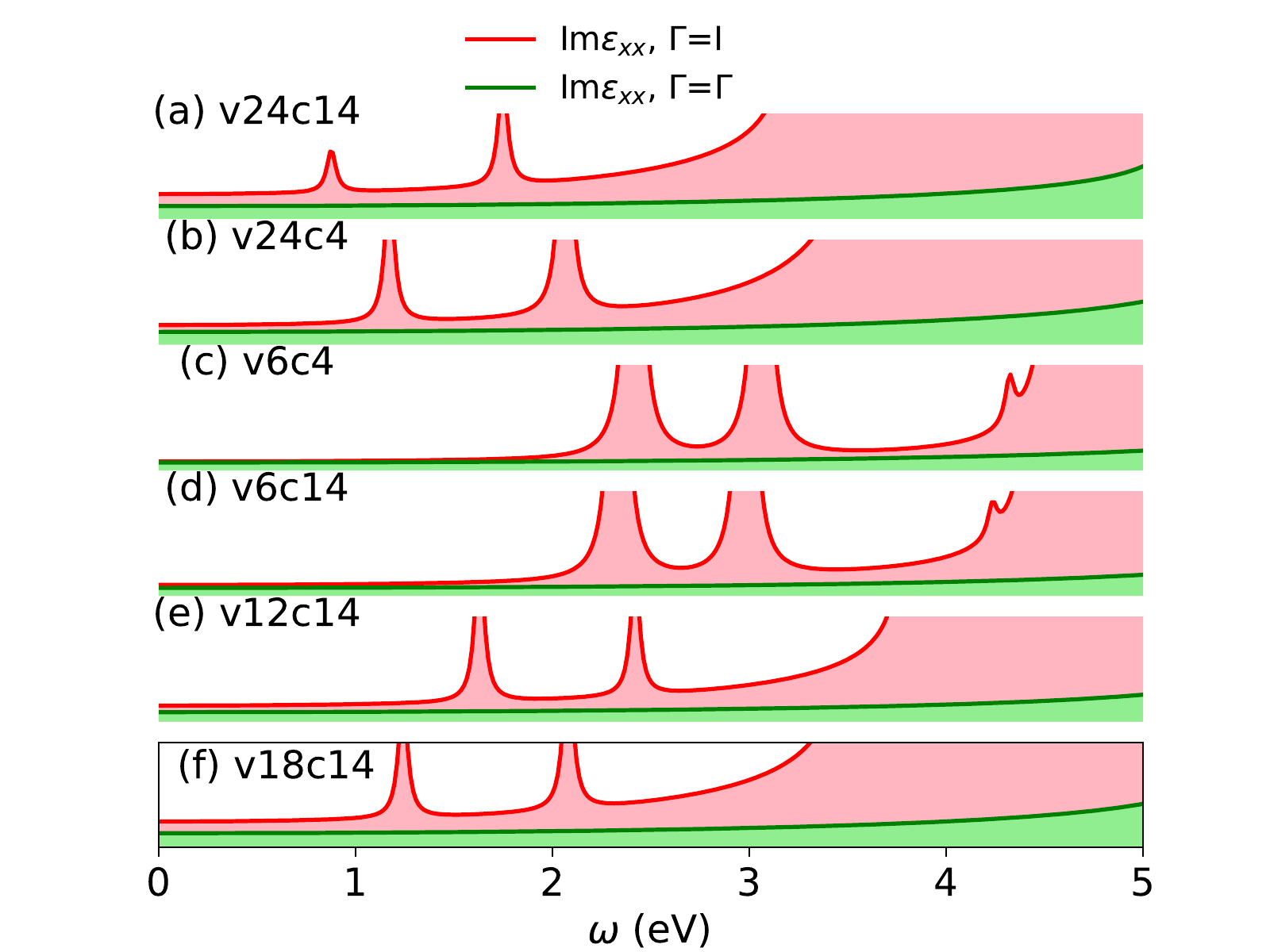}
			\caption{\ce{CrCl3} Monolayer:  Covergence in $E_{ex}$ with the two-particle Hamiltonian size by varying the numbers of valence (v) and conduction bands (c). The green curves that are for the RPA-Im$\epsilon_{xx}$ (the vertex $\Gamma$ put to $I$) also changes for different vc choices, since the one-particle eigenfunctions change in each case. The exciton peaks and the optical spectral weight gets redshifted with larger v-c choices, as more screening channels are included. The $E_{ex}$ is most sensitive to v and weakly depends on $c$ for $c$ $\geq$4 (the crtical number of bands that contain mostly the Cr-$d$-e$_{g\uparrow}$ states).}
			\label{fig:hamilcl}
		\end{center}
\end{figure}

\begin{figure*}
	\begin{center}
\includegraphics[width=1\columnwidth, angle=-0]{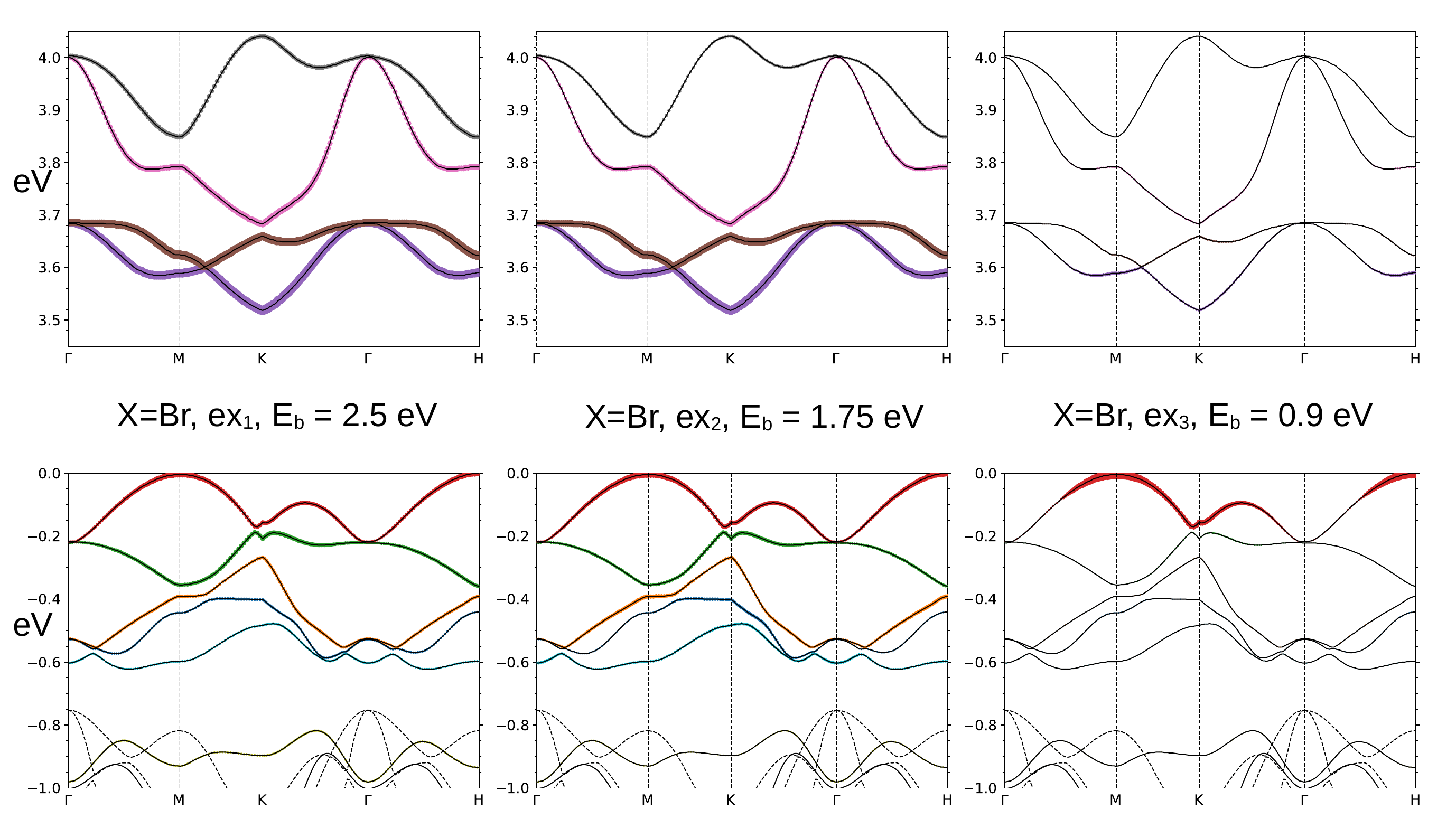}		
		\caption{\ce{CrBr3} bulk : The spectral weight analysis for the three deepest lying exciton ex$_{1}$, ex$_{2}$ and ex$_{3}$ from left to right respectively.  Almost all the bands containing Cr-3$d$ t$_{2g\uparrow}$ and e$_{g\uparrow}$ orbital characters contribute to ex$_{1,2}$. Exciton spectral weight is almost uniformly spread across these bands in case of ex$_{1,2}$. Spectral weight for ex$_{3}$, which has lesser binding energy compared to ex$_{1,2}$, becomes localized in band basis.}
		\label{fig:excitonbr}
	\end{center}
\end{figure*}

\begin{figure*}
	\begin{center}
\includegraphics[width=1\columnwidth, angle=-0]{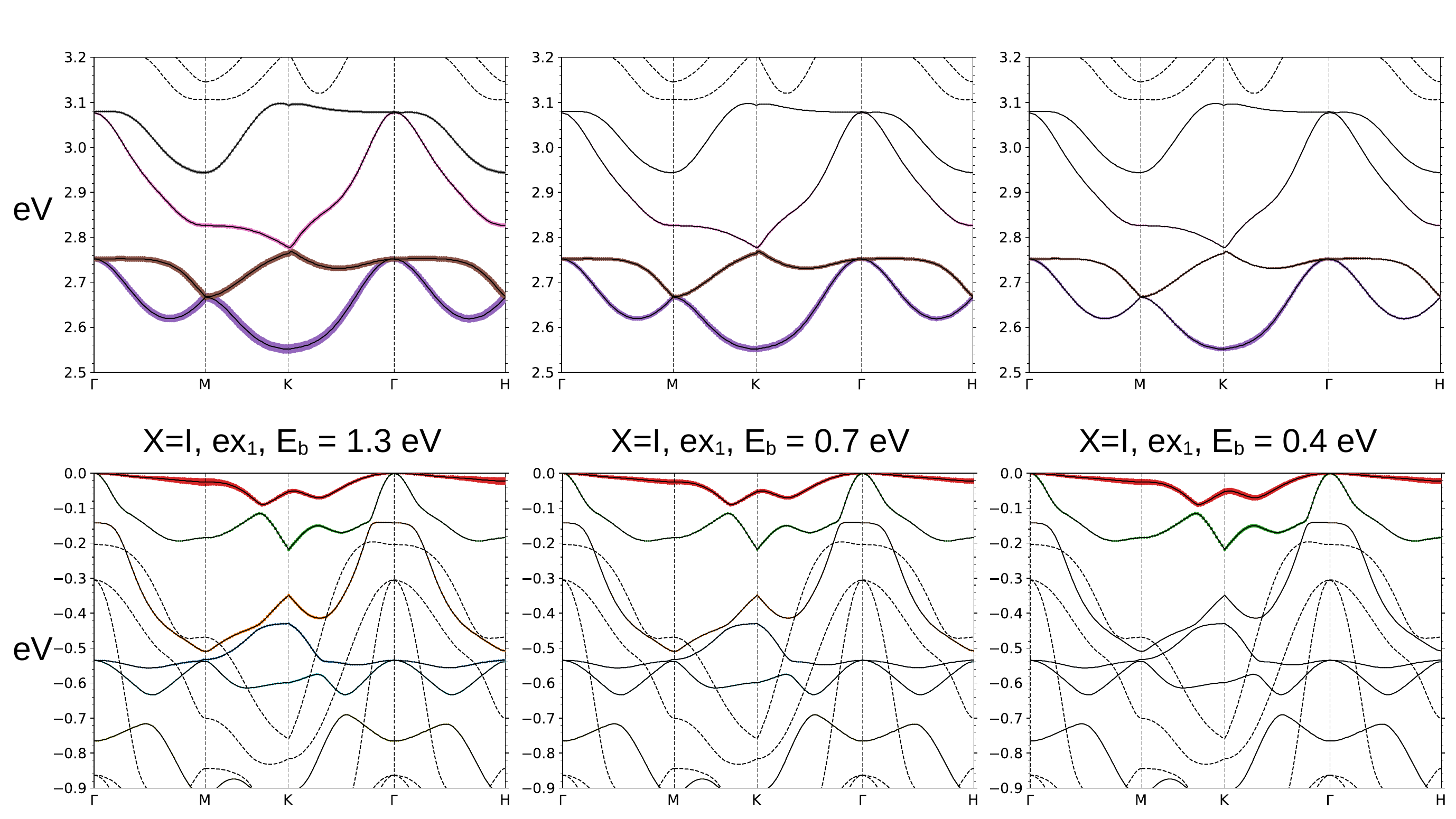}	
	\caption{\ce{CrI3} bulk : The spectral weight analysis for the three deepest lying exciton ex$_{1}$, ex$_{2}$ and ex$_{3}$ from left to right respectively.  Almost all the bands containing Cr-3d t$_{2g\uparrow}$ and e$_{g\uparrow}$ orbital characters contribute to ex$_{1,2}$. Exciton spectral weight is almost uniformly spread across these bands in case of ex$_{1,2}$. Spectral weight for ex$_{3}$, which has lesser binding energy compared to ex$_{1,2}$, becomes localized in band basis. }
		\label{fig:excitonI}
	\end{center}
\end{figure*}


\end{document}